\documentclass{iopjournal}
\usepackage[english]{babel}
\usepackage{todonotes}
\usepackage{lineno}
\usepackage{hyperref}
\renewcommand{\headrulewidth}{0pt}
\usepackage{amsmath}
\usepackage{graphicx}
\usepackage{xcolor}
\usepackage[colorlinks=true, allcolors=blue]{hyperref}
\usepackage{subfigure}
\usepackage{subcaption}
\usepackage[square,numbers,sort]{natbib}
\usepackage{lmodern}

\begin{document}
%\articletype{article}

%\setcounter{totalnumber}{1}
\title{High-rate accumulation of ${8.9\times10^9}$ positrons using a linac-driven buffer-gas trap combined with a  SiC remoderator}

% mydefinitions %%%%%%%%%%%%%%%%%%%%%%%%%%%%%%%%%%%%%%%%%%%%%%%
\newcommand{\uli}{\underline}
\newcommand{\oli}{\overline}
\newcommand{\ftns}{\footnotesize}
\newcommand{\bm}{\boldmath}
\newcommand{\trt}{\ensuremath{^{\mbox{\tiny{\textregistered}}} }}
\newcommand{\noi}{\noindent}
\newcommand{\tbc}{{~\color{red}(to be checked)}}
\newcommand{\tbi}{{~\color{red}(to be improved)}}
\newcommand{\tbf}{{~\color{red}(to be fixed)}}

\newcommand{\alp}{\ensuremath{\alpha}}
\newcommand{\bet}{\ensuremath{\beta}}
\newcommand{\gam}{\ensuremath{\gamma}}
\newcommand{\sig}{\ensuremath{\sigma}}

\newcommand{\pro}{\ensuremath{\mathrm{p}}}
\newcommand{\neu}{\ensuremath{\mathrm{n}}}
\newcommand{\pbar}{\ensuremath{\mathrm{\oli p}}} % \newcommand{\pbar}{$\oli{\text{p}}$\,}
\newcommand{\pbard}{\ensuremath{\mathrm{\oli{p}d}}}
\newcommand{\pbarp}{\ensuremath{\mathrm{p\oli p}}}
\newcommand{\qbar}{\ensuremath{\mathrm{\oli q}}}
\newcommand{\Hbar}{\ensuremath{\mathrm{\oli H}}}% \newcommand{\Hbar}{$\oli{\text{H}}$\,}
\newcommand{\Hbarplus}{\ensuremath{\mathrm{\oli H}^+}}
\newcommand{\Hminus}{\ensuremath{\mathrm{H}^-}}
\newcommand{\Beplus}{\ensuremath{\mathrm{Be}^+}}

\newcommand{\Hneutral}{\ensuremath{\mathrm{H}}}%\newcommand{\Hminus}{$\text{H}^-$\,}

\newcommand{\Ly}{Lyman-$\alpha$\,} 
\newcommand{\LyD}{Ly-$\alpha$D\,}
\newcommand{\pbT}{\pbar\ trap}

\newcommand{\piz}{\ensuremath{\pi^{0}}}
\newcommand{\pip}{\ensuremath{\pi^{+}}}
\newcommand{\pipm}{\ensuremath{\pi^{\pm}}}
\newcommand{\pimp}{\ensuremath{\pi^{\mp}}}
\newcommand{\pim}{\ensuremath{\pi^{-}}}
\newcommand{\gaga}{\ensuremath{\gamma\gamma}}

\newcommand{\kr}{\ensuremath{\,\mathrm{^{83m}Kr}}}
\newcommand{\ar}{\ensuremath{\,\mathrm{^{39}Ar}}}
\newcommand{\co}{\ensuremath{\,\mathrm{^{57}Co}}}
\newcommand{\cf}{\ensuremath{\,\mathrm{^{252}Cf}}}
\newcommand{\ev}{\ensuremath{\,\mathrm{eV}}}
\newcommand{\kev}{\ensuremath{\,\mathrm{keV}}}
\newcommand{\mev}{\ensuremath{\,\mathrm{MeV}}}
\newcommand{\gev}{\ensuremath{\,\mathrm{GeV}}}
\newcommand{\cm}{\ensuremath{\,\mathrm{cm}}}
\newcommand{\mum}{\ensuremath{\,\mu\mathrm{m}}}
\newcommand{\murad}{\ensuremath{\,\mu\mathrm{m}\cdot\mathrm{rad}}}
\newcommand{\cmsq}{\ensuremath{\,\mathrm{cm^{2}}}}
\newcommand{\cmcb}{\ensuremath{\,\mathrm{cm^{3}}}}
\newcommand{\mgcs}{\ensuremath{\,\mathrm{mg/cm^{2}}}}
\newcommand{\mm}{\ensuremath{\,\mathrm{mm}}}
\newcommand{\um}{\ensuremath{\,\mu\mathrm{m}} }
\newcommand{\mmsq}{\ensuremath{\,\mathrm{mm^{2}}}}
\newcommand{\nm}{\ensuremath{\,\mathrm{nm}}}
\newcommand{\mis}{\ensuremath{\,\mathrm{ms}}}
\newcommand{\us}{\ensuremath{\,\mu\mathrm{s}} }
\newcommand{\ns}{\ensuremath{\,\mathrm{ns}}}
\newcommand{\kv}{\ensuremath{\,\mathrm{kV}}}
\newcommand{\ohm}{\ensuremath{\,\Omega}}
\newcommand{\hz}{\ensuremath{\,\mathrm{Hz}}}
\newcommand{\khz}{\ensuremath{\,\mathrm{kHz}}}
\newcommand{\mhz}{\ensuremath{\,\mathrm{MHz}}}
\newcommand{\degr}{\ensuremath{^{\circ}}}

\newcommand{\nclu}{\ensuremath{n_{clu}}}
\newcommand{\Sclu}{\ensuremath{\Sigma_{clu}}}

\newcommand\ainv{A' \to invisible}
\newcommand\ma{m_{A'}}
\newcommand\pair{e^+e^-}
\newcommand\na{{n}_{A'}}
\newcommand\Na{{N}_{A'}}
\newcommand\ea{e^- Z \to e^- Z A'; A' \to invisible}
\newcommand{\en}{\ensuremath{\,\mathrm{e^-}}}
\newcommand{\ep}{\ensuremath{\,\mathrm{e^+}}}
\newcommand{\ps}{\ensuremath{\,\mathrm{Ps}}}
\newcommand\emu{e^- Z \to e^- Z \g; \g \to \mu^+ \mu^-}
\newcommand\dimu{\mu^+ \mu^-}
\newcommand\ee{e^+e^-}
\newcommand\aee{A'\to e^+e^-}
\newcommand\xee{X\to e^+e^-}
\newcommand\Colorhref[3][cyan]{\href{#2}{\color{#1}#3}}
% end mydefinitions %%%%%%%%%%%%%%%%%%%%%%%%%%%%%%%%%%%%%%%%%%%%%%%

\author{
P~Adrich$^{1}$\orcid{0000-0002-7019-5451},
I~Belosevic$^{2}$\orcid{0000-0002-8001-8889},
M~Chung$^{3}$\orcid{0000-0001-7014-4120},
P~Cladé$^{4}$\orcid{0000-0001-5070-8734},
P~Comini$^{2}$\orcid{0000-0002-6373-4752},
P~Crivelli$^{5}$\orcid{0000-0001-5430-9394},
P~Debu$^{2}$\orcid{0000-0003-2988-5052},
A~Douillet$^{4,6}$\orcid{0000-0001-6927-6775},
S~Geffroy$^{7}$\orcid{0009-0009-6940-8120},
S~Guellati-Khelifa$^{4}$\orcid{0000-0002-8412-411X},
P~Guichard$^{8}$\orcid{0009-0003-8568-5327},
P-A~Hervieux$^{8}$\orcid{0000-0002-4965-9709},
L~Hilico$^{4,6}$\orcid{0000-0002-8916-1294},
P~Indelicato$^{4}$\orcid{0000-0003-4668-8958},
S~Jonsell$^{9}$\orcid{0000-0003-4969-1714},
J-P~Karr$^{4,6}$\orcid{0000-0003-2082-0914},
B~Kim$^{10}$\orcid{0000-0003-0958-5503},
S~Kim$^{11}$\orcid{0000-0002-0013-0775},
E-S~Kim$^{12}$\orcid{0000-0001-5603-6764},
N~Kuroda$^{13}$\orcid{0000-0003-2727-790X},
B~Lee$^{11}$\footnote{Present address: QUANTUM, Institut für Physik, Johannes Gutenberg Universität, D-55128 Mainz, Germany}\orcid{0000-0002-7293-5142},
L~Liszkay$^{2,*}$\orcid{0000-0003-4371-4380},
D~Lunney$^{7}$\orcid{0000-0002-3227-305X},
G~Manfredi$^{8}$\orcid{0000-0002-5214-8707},
B~Mansoulié$^{2}$\orcid{0000-0001-5945-5518},
V~Martimort$^{14}$\orcid{0009-0005-6389-8465},
M~Matusiak$^{1}$\orcid{0000-0002-8239-6971},
S~Niang$^{2}$\orcid{0000-0002-0481-252X},
V~Nesvizhevsky$^{15}$\orcid{0000-0002-5364-0197},
F~Nez$^{4}$\orcid{0000-0002-3478-7521},
K~Park$^{11}$\orcid{0000-0002-6013-0259},
N~Paul$^{4}$\orcid{0000-0003-4469-780X},
E~Perez$^{16}$\orcid{0000-0002-7359-9689},
P~Pérez$^{2,17}$\orcid{0000-0003-1407-1582},
C~Regenfus$^{5}$\orcid{0000-0001-9656-3104},
C~Roumegou$^{7}$\orcid{0009-0006-0343-9256},
J-Y~Roussé$^{2}$,
Y~Sacquin$^{2}$,
F~Schmidt-Kaler$^{18}$\orcid{0000-0002-5697-2568},
K~Szymczyk$^{1}$\orcid{0000-0001-9159-485X},
T~A~Tanaka$^{13}$\footnote{Present Address: National Metrology Institute of Japan (NMIJ), National Institute of Advanced Industrial Science and Technology (AIST), 1-1-1 Umezono, Tsukuba, Ibaraki 305-8568, Japan}\orcid{0009-0003-8306-5546},
B~Tuchming$^{2}$\orcid{0000-0002-1356-0723},
B~Vallage$^{2}$\orcid{0000-0003-1255-8506},
D~P~van~der~Werf$^{17}$\orcid{0000-0001-5436-5214},
D~Won$^{11}$\orcid{0009-0005-5557-7709},
S~Wronka$^{1}$\orcid{0000-0003-3277-138X} and
P~Yzombard$^{4}$\orcid{0000-0002-0864-181X}
(GBAR Collaboration)
}

\affil{
$^{1}$ National Centre for Nuclear Research (NCBJ), ul. Andrzeja Soltana 7, 05-400 Otwock, Swierk, Poland\\
$^{2}$ IRFU, CEA, Université Paris-Saclay, F-91191 Gif-sur-Yvette, France\\
$^{3}$ The Pohang University of Science and Technology (POSTECH), Pohang, Gyeongbuk province, Chungam-Ro77 Hyoja-Dong, Nam-Gu, Pohang, Kyungbuk, 37673, Republic of Korea\\
$^{4}$ Laboratoire Kastler Brossel, Sorbonne Université, CNRS, ENS-Université PSL, Collège de France, Campus Pierre et Marie Curie, 4, place Jussieu, F-75005 Paris, France\\
$^{5}$ Institute for Particle Physics and Astrophysics, ETH Zurich, 8093 Zurich, Switzerland\\
$^{6}$ Université d'Evry-Val d'Essonne, Université Paris-Saclay, Boulevard François Mitterand, F-91000 Evry, France\\
$^{7}$ Université Paris-Saclay, CNRS/IN2P3, IJCLab, Orsay, France\\
$^{8}$ Université de Strasbourg, CNRS, Institut de Physique et Chimie des Matériaux de Strasbourg, UMR 7504, F-67000 Strasbourg, France\\
$^{9}$ Department of Physics, Stockholm University, SE-10691 Stockholm, Sweden\\
$^{10}$ Center for Underground Physics, Institute for  Basic Science, 70 Yuseong-daero 1689-gil,  Yuseong-gu, Daejeon 34047, Korea\\
$^{11}$ Department of Physics and Astronomy, Seoul National University, 1 Gwanak-Ro, Gwanak-gu, Seoul 08826, Korea\\
$^{12}$ Department of Accelerator Science, Korea University Sejong Campus, Sejong-ro 2511, 0019 Sejong, Republic of Korea\\
$^{13}$ Institute of Physics, University of Tokyo, 3-8-1 Komaba, Meguro, Tokyo 153-8902, Japan\\
$^{14}$ Laboratoire Matériaux et Phénomènes Quantiques, Université Paris Cité, CNRS UMR 7162, 75013 Paris, France\\
$^{15}$ Institut Max von Laue - Paul Langevin (ILL), 71 avenue des Martyrs, F-38042 Grenoble, France\\
$^{16}$ CERN, Esplanade des Particules 1, 1217 Meyrin, Switzerland\\
$^{17}$ Department of Physics, Faculty of Science and Engineering, Swansea University, Swansea SA2 8PP, United Kingdom\\
$^{18}$ QUANTUM, Institut für Physik, Johannes Gutenberg Universität, D-55128 Mainz, Germany\\
$^{*}$ Author to whom any correspondence should be addressed.} 

\email{laszlo.liszkay@cea.fr}


\begin{abstract}
We report the accumulation of $(8.9 \pm 0.9)\times10^9$ positrons in less than 35 minutes in a Penning--Malmberg trap connected to the GBAR electron-linac-based positron source. To our knowledge, this exceeds the largest previously reported trapped-positron population by more than a factor of two and was achieved at an 18-fold higher average accumulation rate of $4.5\times10^6$~e$^+$/s. The accumulated positron clouds are intended for the creation of dense positronium targets to produce keV-antihydrogen beams by charge exchange with antiprotons delivered by the CERN AD/ELENA facility. The system combines a low-energy electron linac, a modified Surko-type buffer-gas trap enhanced by a silicon carbide remoderator, and a high-field Penning--Malmberg trap operated as a positron accumulator. Time-resolved annihilation measurements provide insight into the capture and cooling dynamics, while comparisons of CO$_2$ and SF$_6$ show that the trapping signal reaches saturation at lower gas pressures with SF$_6$. Measurements show that the SiC-remoderator-based trapping scheme more than doubles the trapping efficiency, contributing to a more than threefold increase in the positron accumulation rate compared with the original configuration.
\end{abstract}
\section{Introduction}

Experiments with low-energy antimatter provide sensitive tests of the Standard Model of particle physics through tests of CPT symmetry as well as the means to verify the gravitational interaction of antimatter \cite{Charlton_2020_book}. CERN's Antiproton Decelerator and Extra Low ENergy Antiproton facility (AD/ELENA), unique worldwide, provides antiproton beams decelerated to 100~keV and hosts several experiments performing or preparing precision tests of fundamental symmetries \cite{ELENA}. The ALPHA \cite{akbari_alpha-2_2025}, ASACUSA \cite{imao_positron_2010}, AEgIS \cite{Kellerbauer_2008}, and GBAR \cite{perez_2015} experiments in the AD/ELENA hall produce and study antihydrogen, the bound state of an antiproton and a positron. In these experiments, with the exception of GBAR, the low-energy positrons required for antihydrogen production are obtained using the well-established techniques of moderation and accumulation from a radioactive $^{22}$Na source, pioneered by Surko et al. \cite{Surko1989}.

%GBAR has chosen to build a positron generator which is based on a low energy (9 MeV) linac, dedicated to the experiment. This device provides a higher positron flux than standard isotope-based sources and it can be switched off without significant remaining radioactivity. It fills a set of particle traps which eventually allows generating an intense positron pulse.

The GBAR experiment is designed to test the weak equivalence principle by measuring the free-fall acceleration of antihydrogen in the Earth's gravitational field \cite{perez_2015}. Recent measurements by the ALPHA-g collaboration \cite{alpha-g_result_2023} showed that the gravitational behaviour of antihydrogen is compatible with that of ordinary matter and ruled out a purely repulsive gravitational interaction, although the experimental uncertainty remained relatively large. In its first phase, GBAR aims to determine the gravitational acceleration of antihydrogen in an environment free of electromagnetic fields with a relative precision of approximately 1~\%, with substantially improved precision envisaged in a later phase. To achieve this goal, GBAR plans to produce antihydrogen ions that can be sympathetically cooled before photodetachment and free-fall measurement. The ions are intended to be produced through two successive charge-exchange reactions with a dense positronium (Ps) cloud:
\begin{equation}
    \overline{\mathrm{p}} + \mathrm{Ps}
    \rightarrow
    \overline{\mathrm{H}} + \mathrm{e}^{-},
    \label{eq1}
\end{equation}

\begin{equation}
    \overline{\mathrm{H}} + \mathrm{Ps}
    \rightarrow
    \overline{\mathrm{H}}^{+} + \mathrm{e}^{-}.
    \label{eq2}
\end{equation}
Here, \pbar\ denotes an antiproton and \Hbar\ antihydrogen. GBAR has successfully demonstrated the production of \Hbar\ through the first of these reactions \cite{Adrich_2023}. Producing \Hbarplus through these two reactions requires a high positronium density and therefore far more positrons than can practically be supplied by conventional $^{22}$Na sources. To overcome this limitation, the GBAR collaboration has developed a high-flux positron source based on a low-energy (9~MeV) electron linear accelerator \cite{Charlton_2021}. In the first version of the installation, the positrons produced by the source were moderated, cooled in a Surko-type buffer-gas trap, and subsequently accumulated in a Penning--Malmberg trap at a rate of $1.3\times10^{6}$ e$^+$/s, reaching 1.4 billion in 18 minutes \cite{Blumer_2022}. The apparatus described in the present work is an improved version of the GBAR positron beamline.

The principal improvement over the system described in Ref.~\cite{Blumer_2022} is a modified buffer-gas-trap configuration incorporating a silicon carbide (SiC) remoderator. This scheme was first demonstrated by us in the GBAR positron beamline in 2021 \cite{liszkay_2021,Blumer_2022a} and has been used in the production setup since 2023. A similar approach was subsequently reported by Michishio et al.~\cite{Michishio_2022}. In the new scheme the conventional nitrogen-gas-based capture stage is replaced by implantation following remoderation in a single-crystal silicon-carbide (SiC) wafer. This modification \cite{Liszkay_2025}, together with the optimization of the transfer of particles to the high-field trap, increased the accumulation rate by a factor of approximately 3.5, from $(1.3\pm0.2)\times10^{6}$~e$^+$/s to $(4.5\pm0.5)\times10^{6}$~e$^+$/s. These results suggest that SiC-remoderator-based buffer-gas traps are better adapted to linac-based positron sources than the conventional Surko scheme \cite{Blumer_2022,Higaki_2020}. The reason may be that the energy distribution of linac beams is less suitable for capture by collision with nitrogen molecules, although the experimental evidence is not sufficient for a definite conclusion. Similar installations also have the potential to supply high-quality positron beams with narrow energy distribution and small diameter \cite{weber_creation_2008,natisin_formation_2015,natisin_cryogenically_2016} for depth-dependent positron-annihilation spectroscopy \cite{tuomisto_defect_2013}, surface studies using positrons \cite{Hugenschmidt_2008} and high-resolution investigations of positron-gas interaction \cite{gribakin_positron-molecule_2010}.   

This work describes the improved positron accumulation scheme, which has enabled the accumulation of $(8.9\pm0.9)\times10^{9}$ positrons, double the previously published highest value \cite{Fitzakerley_2016}, with an approximately 18-times higher accumulation rate. The GBAR positron beamline is described in Sec.~\ref{sec:exp} and the operation of its principal components and the results are discussed in Sec.~\ref{sec:discussion}.

\section{Experimental setup}
\label{sec:exp}
\begin{figure}
\begin{center}
\includegraphics[height=3cm]{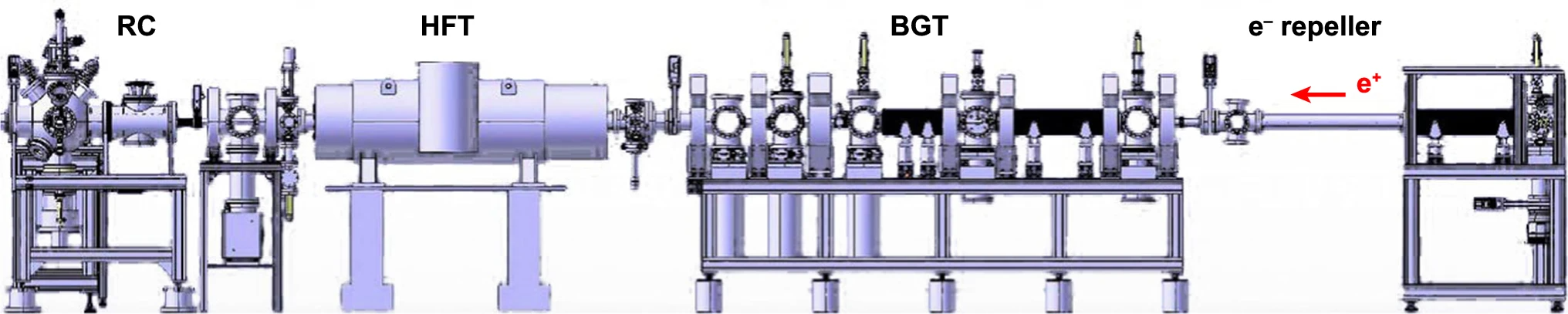}
\caption{The GBAR positron transport line:  positrons arrive from the linac (right, not shown), are cooled in the buffer-gas trap (BGT), accumulated in the high-field trap (HFT) and sent to the reaction chamber target (RC) for producing positronium.  The antiproton beam traverses the reaction chamber perpendicular to the plane of the figure.}
\label{fig:eplusline}
\end{center}
\end{figure}

A schematic layout of the entire GBAR experiment, together with brief descriptions of its main components, can be found in Fig.~1 of Ref. \cite{Adrich_2023}. Here, Fig.~\ref{fig:eplusline} shows a schematic representation of the positron beamline.

\subsection{The linac-based positron source}
\label{sec:linac}
The positron source is based on a compact, water-cooled linear accelerator, operating at an electron energy of 9 MeV and a peak current of 300~mA \cite{Charlton_2021}. The accelerator delivers electron pulses of 2.5~\us duration at repetition rates up to 300~Hz; during routine operation the frequency is limited to 200~Hz to extend system lifetime. The electron beam impinges on a 1-mm-thick, water-cooled tungsten target, where bremsstrahlung photons generate electron–positron pairs in the target and in a downstream moderator.

The moderator consists of a stack of approximately 18 layers of thin tungsten mesh (20~\um wire thickness), annealed at temperatures above 2400~\degr C under vacuum in a separate chamber by Joule heating. Emitted positrons are extracted and accelerated to 50–150 eV using a grounded extraction electrode with a central aperture. A time-dependent bunching potential is applied in the present setup to reduce the positron pulse width to approximately 1.5~\us at the entrance of the buffer-gas trap. Shorter positron pulses reduce the fraction of positrons that return to the remoderator surface and annihilate.

Positrons are transported through a 7.5-m beamline by an axial magnetic field of 8~mT, passing through the linac biological shielding and incorporating two 90\degr~bends to limit radiation outside the bunker. Co-transported electrons from the target region are suppressed by an ``electron repeller'', a 5~kV electrostatic barrier formed by tubular electrodes located outside the bunker \cite{Blumer_2022}.

\subsection{The buffer-gas trap (BGT)}
\label{sec:buffer}
\begin{figure}
	\centering
	\begin{subfigure}[]
		\centering
         	\includegraphics[width=0.45\linewidth]{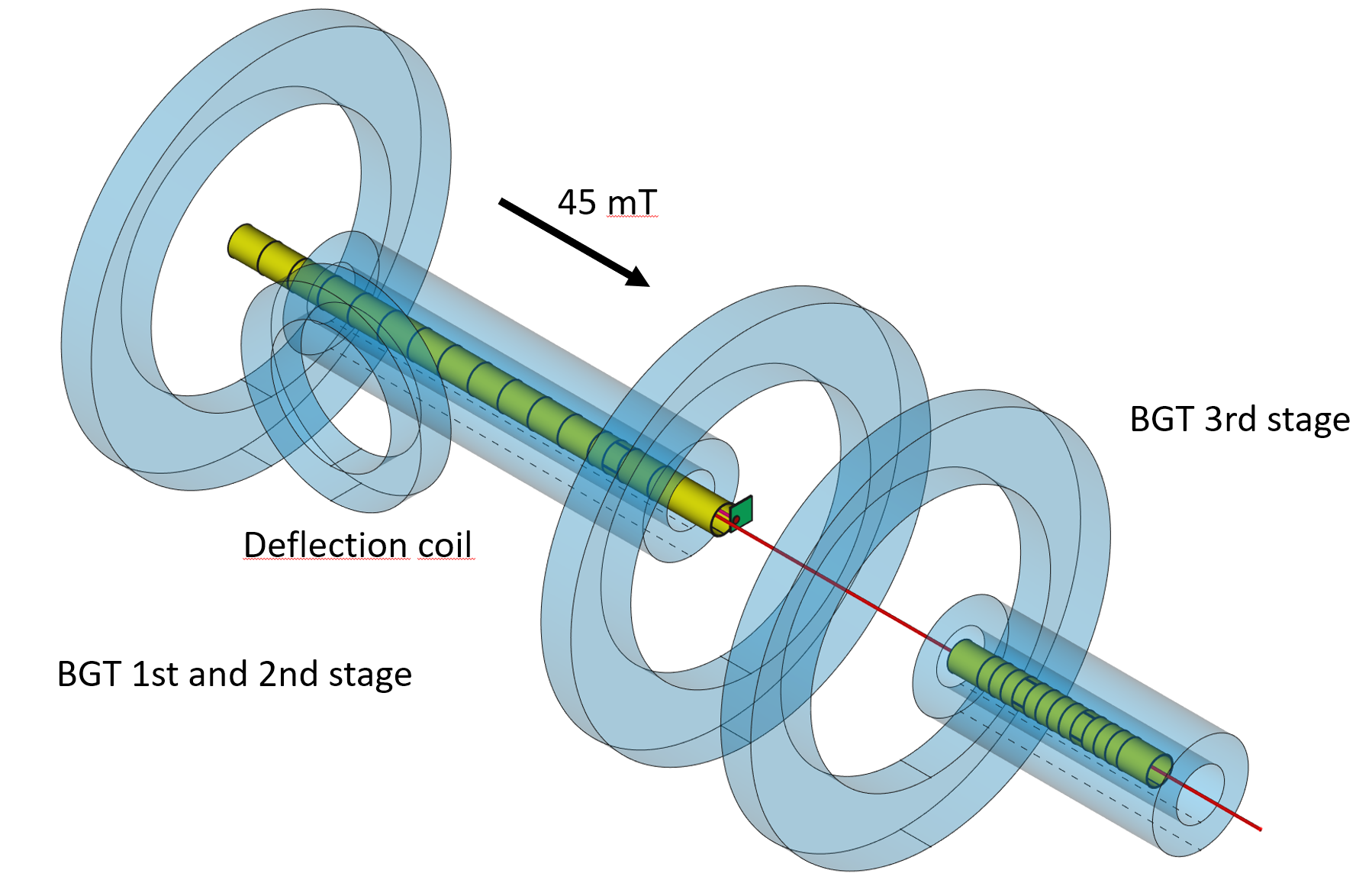}
        \label{fig:BGT_a}
	\end{subfigure}
	\begin{subfigure}[]
		\centering
         	\includegraphics[width=0.45\linewidth]{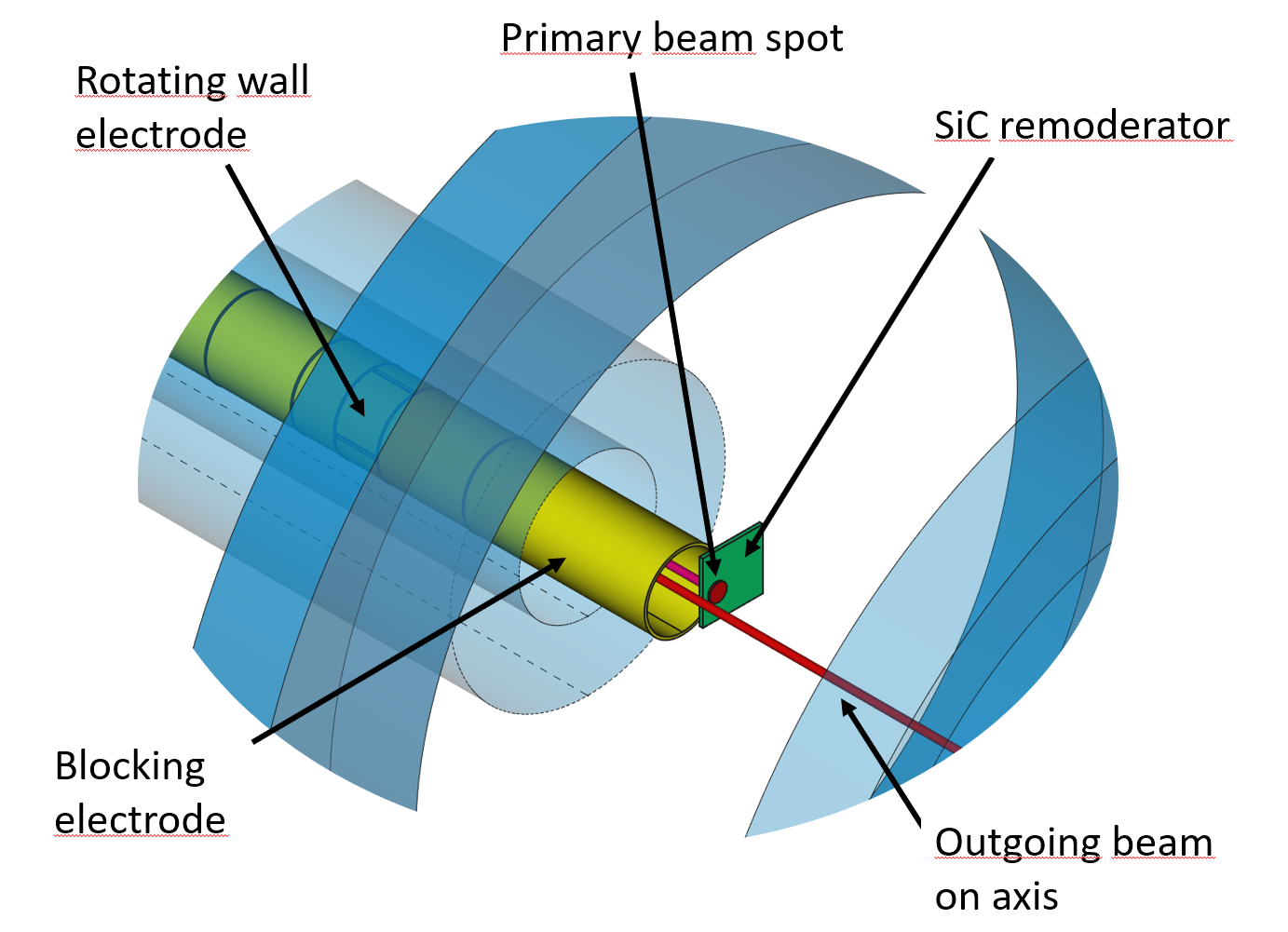}
        \label{fig:BGT_b}
	\end{subfigure}
     	\caption{(a) Schematic drawing of the buffer-gas trap electrodes and coils; (b) an enlarged view of the positron remoderator. (Reproduced from Ref. \cite{Liszkay_2025}.) }
     	\label{fig:BGT}
\end{figure}

Positrons are captured in a Penning–Malmberg trap formed by the first electrode stack of the buffer-gas trap (Fig.~\ref{fig:BGT}(a)). The electrode assembly previously described in \cite{Blumer_2022} has been replaced by a redesigned stack optimized for operation with a silicon carbide remoderator–based trapping scheme. Electrodes with reduced inner diameter, originally required for nitrogen-buffer-gas capture, were removed and replaced by larger-diameter electrodes, and the overall length of the stack was increased. The capture section is 665~mm long and consists of eleven 50-mm electrodes and one 105-mm electrode, all with an inner diameter of 41~mm. One electrode is segmented into four azimuthal sectors to apply rotating-wall compression using phase-shifted sinusoidal signals capacitively coupled to the DC bias. The electrodes are driven by fast high-voltage amplifiers providing potentials up to ±140~V \cite{Samuel_thesis}. Trap operation, including capture, cooling, and ejection, is controlled by an FPGA-based sequencer, with particle extraction performed using a fast high-voltage switch. Carbon dioxide is used as the cooling gas and is injected at the downstream end of the stack, establishing pressures of approximately  $4\times10^{-6}$~mbar upstream and $1.6\times10^{-4}$~mbar downstream.

The incoming positron beam is deflected by a magnetic dipole field generated by a steering coil positioned upstream of the first trapping stage, directing the particles onto an off-axis remoderator, a 20~mm$\times$20~mm SiC plate, downstream of the capture region (shown in Fig.~\ref{fig:BGT}(a)). Operated in reflection mode, it re-emits positrons at low energy with an efficiency exceeding 50~\% \cite{Leite_2017}. The remoderator consists of an n-type nitrogen-doped 4H-SiC (0001) single-crystal wafer with a resistivity of 16--28~m$\Omega\cdot$cm, mounted on a linear translator. It can be moved onto the beam axis where it blocks positrons ejected from the second stage and allows measurement of the trapping efficiency, or moved off axis to allow transfer of particles to the subsequent trapping stage. Its bias is supplied by an amplifier of the same type as for the electrodes, with fast potential modulation applied via capacitive coupling to a high-frequency arbitrary waveform generator. 

The third trapping stage of the BGT comprises two 49-mm and twelve 18.4-mm-long electrodes with 41~mm inner diameter, one of which is segmented into four sectors for rotating-wall compression. The CO$_2$ pressure in this region is approximately 
$4\times10^{-7}$ mbar. Water-cooled solenoids provide an axial magnetic field of 45~mT along the electrode stacks.

\subsection{The high-field trap (HFT)}
\label{sec:hft}

\begin{figure}
\begin{center}
\includegraphics[height=6cm]{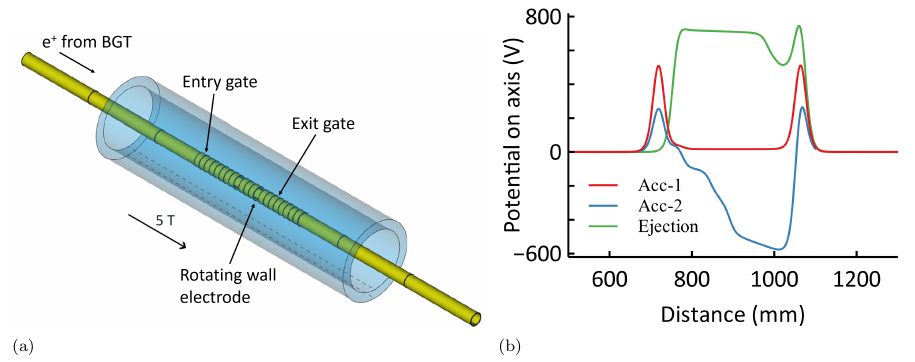}
\caption{(a) Schematic drawing of the high-field trap. (b) Axial potential profiles at the beginning (red) and end (blue) of the accumulation of $8.9\times10^9$ positrons, and immediately before ejection (green).}
\label{fig:trap}
\end{center}
\end{figure} 

The HFT was developed at RIKEN \cite{Oshima_2003}, where electron densities of up to $10^{10}$ cm$^{-3}$ were achieved.  It was later adapted to capture positrons from a low-energy electron linac test bench in Saclay \cite{Grandemange, Lunney2014}. The HFT is a Penning-Malmberg trap operated in a 5-T magnetic field generated by a superconducting solenoid. The electrode stack (Fig.~\ref{fig:trap}(a)) is positioned at the center of the 500-mm-long section of the trap, where the field homogeneity is better than 10$^{-3}$ within a 2~mm radius. The central section comprises 23-mm-long cylindrical electrodes, including two half-length electrodes segmented into four sectors for rotating-wall excitation. Electrode inner surfaces are coated with colloidal graphite to suppress patch-potential effects.

Trapping and transport potentials of up to ~2~kV are supplied by high-voltage amplifiers (Ultravolt HVA series, with an RC filter with 160~Hz cutoff frequency), while fast push–pull switches (Behlke) enable rapid potential switching at the trap entrance and exit. The electrode assembly is installed within the 100-mm-diameter magnet bore, cooled to 12 K using a closed-cycle cryocooler. The vacuum pressure in the bore is not directly measurable and is estimated to be on the order of 10$^{-10}$~mbar. Pumping of residual hydrogen gas is enhanced by activated carbon granules cooled by the magnet bore.

\subsection{Detectors}
\label{sec:detectors}
Annihilation $\gamma$ rays are monitored at several locations along the apparatus using caesium iodide (CsI) scintillation detectors coupled to PIN photodiodes. The response of the CsI detectors was calibrated by independently measuring the charge delivered by a positron pulse to a conductive target (either a metallic plate or the front surface of a microchannel plate (MCP) detector). The target was biased at $+100$~V to suppress the emission of secondary electrons, which would otherwise distort the charge measurement. The charge deposited by each positron pulse was measured using a calibrated charge-sensitive preamplifier (FastComTec), providing a measurement of the number of incident positrons. The amplitude of the CsI detector signal was then correlated with the measured charge to verify the linearity of the detector response and to determine the conversion factor between the CsI signal and the number of positrons. 

A plastic scintillator coupled to a fast photomultiplier tube is positioned approximately 500~mm from the remoderator to monitor positron annihilations occurring during the capture and cooling sequence. The detector signal is recorded using a digital oscilloscope with a bandwidth of 500~MHz (LeCroy). Waveforms acquired for 800 linac pulses were analysed to determine the number of detected annihilation events in successive 0.01~ms time bins. This provides the annihilation rate as a function of the time elapsed after the arrival of a positron pulse from the linac source and, consequently, the time dependence of positron losses during the capture and cooling cycle.

Beam profiles downstream of the second stage of the buffer-gas trap and at the exit of the high-field trap are recorded using MCP detectors equipped with phosphor screens and imaged with CMOS cameras.

\section{Experimental results and discussion}
\label{sec:discussion}
\subsection{Capture and cooling in the buffer-gas trap}
\label{sec:capture}
\begin{figure}
	\centering
	\begin{subfigure}[]
		\centering
         	\includegraphics[width=0.45\linewidth]{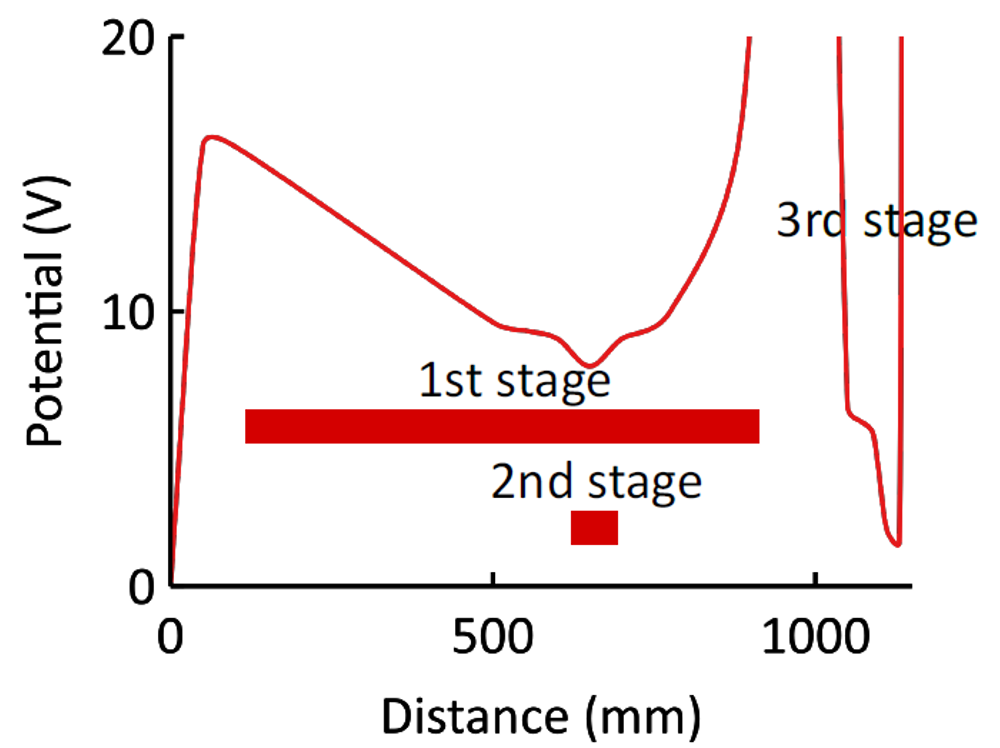}
	\end{subfigure}
	\begin{subfigure}[]
		\centering
         	\includegraphics[width=0.45\linewidth]{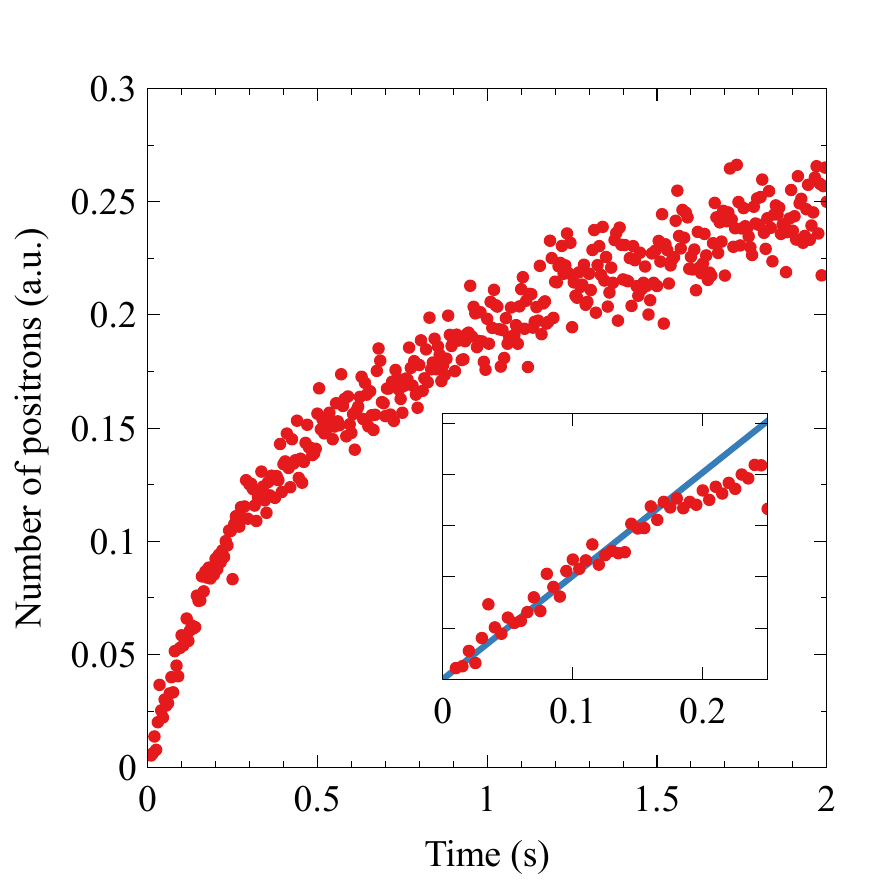}
         	\label{fig:BGT_plots_accumulation}
	\end{subfigure}
     	\caption{Potential along the BGT axis during accumulation (a) and accumulation of positrons in the second stage of BGT (b). The inset shows the short-term accumulation curve with a linear fit using points up to 0.17~s. (Adapted from Ref. \cite{Liszkay_2025}.) }
     	\label{fig:BGT_plots}
\end{figure}
Upon impact on the SiC remoderator, positrons are re-emitted in the backward direction with low kinetic energy, defined by the positron work function of the material \cite{Leite_2017} and the small potential difference applied between the remoderator and the end of the electrode stack at the moment the positron pulse arrives. The re-emission energy is kept low to minimize the number of particles reaching the remoderator after reflection from the opposite end of the trapping potential well before the end of the positron pulse. For the same purpose, the temporal width of the positron pulse is compressed by applying a bunching voltage to the primary moderator. Between positron pulses, the remoderator is biased to a higher potential to prevent stored particles from returning to the surface and annihilating. The field blocks positrons not only at the  moderator but also near the axis of the trap, preventing their escape from the first stage of the trapping sequence.

Following capture, positrons are initially confined in an approximately 600-mm-long potential well forming the first trapping stage, where rapid cooling occurs through collisions with carbon-dioxide gas. Once their axial energy falls below about 1 eV, the particles accumulate in a shorter downstream region of the same trap, constituting the second stage. This region, where rotating-wall compression is continuously applied at 0.5~MHz with an amplitude of 1.4~V$_{\mathrm{pp}}$, is located immediately upstream of the exit barrier (Fig.~\ref{fig:BGT_plots}(a)). Figure \ref{fig:BGT_plots}(b) shows the number of positrons accumulated in the second stage as a function of the accumulation time. The measurement was performed by ejecting the collected particles onto the remoderator, which was moved to the axis of the trap for this experiment.  The curve shows efficient positron accumulation for more than 300~ms.

\begin{figure}
\begin{center}
\includegraphics[width=0.6\linewidth]{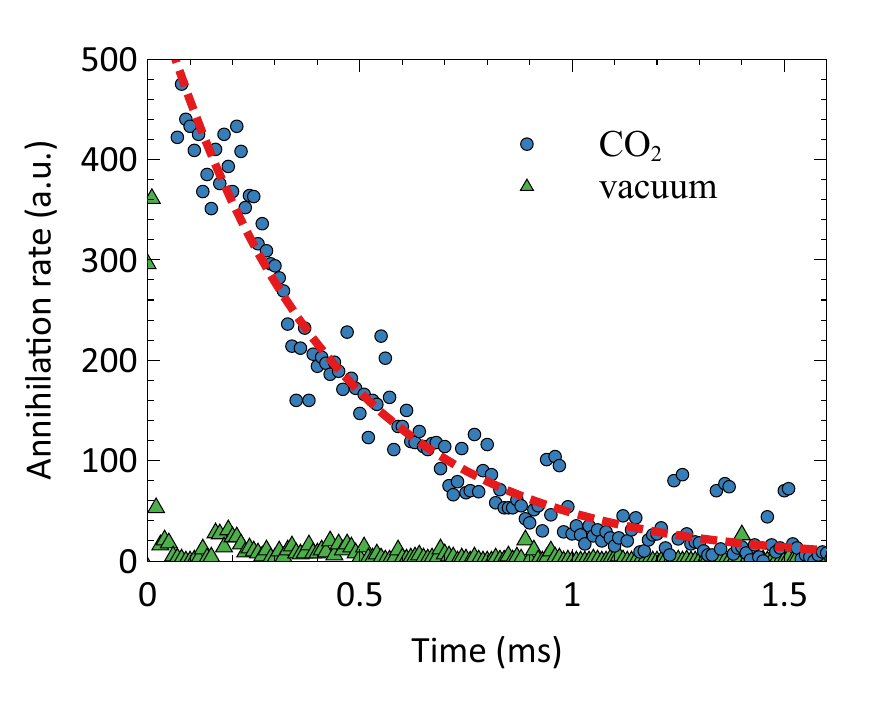}
\caption{Annihilation rate in the first two stages of the buffer-gas trap, measured with a fast scintillation detector, with and without the presence of carbon-dioxide buffer gas. Values below 0.1~ms are distorted by signal pile-up. The dashed red line represents an exponential fit with a 0.4~ms decay constant. }
\label{fig:bgt_loss}
\end{center}
\end{figure} 

The positron cloud is centred on the trap axis by the rotating wall drive and can be transferred to the third trapping stage without intercepting the remoderator, which is positioned off-axis. The feasibility of such transport has previously been demonstrated by Michishio et al. \cite{Michishio_2022} using a remoderator featuring a central aperture. Key operational parameters—including the timing and waveform of the remoderator and bunching voltages, the axial trapping potential profile, and the number of injected positron pulses—were optimized to enhance capture efficiency in the first two stages of the buffer-gas trap.

\begin{figure}
\begin{center}
\includegraphics[width=0.6\linewidth]{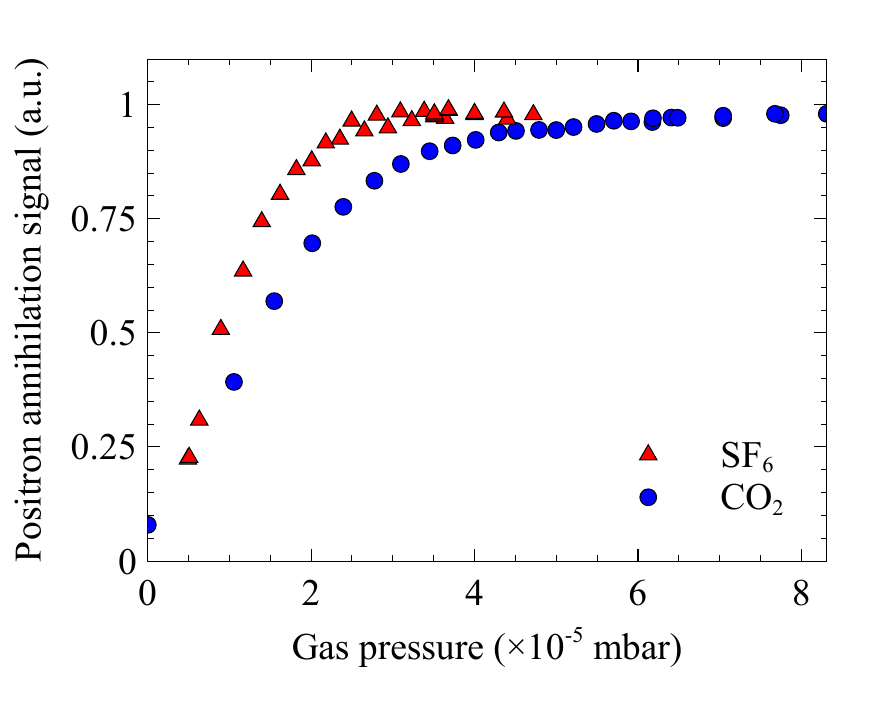}
\caption{Trapping in the buffer-gas trap as a function of gas pressure at the end of the second stage, using carbon dioxide and sulphur hexafluoride. The annihilation signal was recorded after ejecting the particles onto the remoderator, which served as a target. The gas correction factor for the ionization vacuum gauge has been taken into account (1.42 and 2.50 for CO$_2$ and SF$_6$, respectively). The two curves were normalized to the same integrated annihilation signal.}
\label{fig:bgt_gases}
\end{center}
\end{figure}

The time dependence of positron losses during capture and cooling can be investigated using the annihilation signal recorded by the fast detector (Fig.~\ref{fig:bgt_loss}). In the absence of buffer gas, the annihilation signal rapidly drops to nearly zero, consistent with the observation that no detectable positron accumulation occurs. With buffer gas present, the signal, which reflects positron losses through annihilation in the gas or at the trap electrodes, decreases to nearly the background level within approximately 1.5~ms. This suggests that the surviving positrons accumulate in the second stage on a timescale much shorter than the linac repetition period. Measurements performed at higher linac repetition rates show no measurable losses, confirming that carbon-dioxide cooling is sufficiently effective to reduce the positron energy between successive injections and thereby enable confinement in the relatively shallow potential well of the second stage. At the same time, the positron lifetime in this region is long enough to permit accumulation over several hundred milliseconds without significant losses (see Fig.~\ref{fig:BGT_plots}(b)). Carbon dioxide was selected as the cooling gas because it is non-flammable and environmentally benign. Methane has been shown to provide positron cooling times comparable to those obtained with carbon dioxide \cite{Michishio_2022} and is therefore expected to deliver similar performance in the present system. We also compared the capture and cooling efficiency obtained with SF$_6$ (Fig.~\ref{fig:bgt_gases}) and found that the trapping signal reaches saturation at lower gas pressures with SF$_6$. This observation is consistent with the fact that the measured positron cooling time in SF$_6$ is less than one third of that in methane and carbon dioxide \cite{Danielson_2015}.

While direct transfer of positrons from the second trapping stage to the high-field trap (i.e., bypassing the third stage) is in principle possible, this would require additional manipulation of the positron cloud in the cooling gas prior to ejection. Because the duration of such manipulation exceeds the linac repetition period, it would lead to a reduced accumulation duty cycle. For this reason, the third stage of the original buffer-gas trap configuration was retained. In this stage, positrons undergo further cooling in a lower-pressure gas environment (approximately $4\times10^{-7}$ mbar) while being radially compressed using a rotating-wall drive (4.5~MHz, 10~V$_{\mathrm{pp}}$).

During preparation of the positron bunch in the third stage, accumulation in the first two stages proceeds in parallel, minimizing dead time associated with cloud manipulation. The use of a single trapped bunch also simplifies third-stage operation and improves control of the resulting positron cloud. An optimal accumulation duration of 125 ms was identified, balancing storage efficiency with effective transfer to the high-field trap. The combined efficiency of the first two trapping stages is derived by comparing the annihilation signal on the SiC remoderator, moved onto the beam axis and held at a potential of -100 ~V, for a single linac pulse and for the pulse of positrons ejected after accumulation. It is approximately 27~\%, with a transport efficiency of about 85~\% between the second and third stages, estimated from the annihilation signal at the pumping restriction between the two stages. Trapping, compression and ejection in the third stage were achieved without measurable losses. Overall, the SiC-remoderator enhancement achieves more than twice the efficiency of the original Surko-trap configuration \cite{Blumer_2022} when operated with the linac-based positron source.

\subsection{Stacking in the high-field trap}
\label{stacking}
Positrons are extracted from the third stage of the buffer-gas trap after raising the entire potential well, and are transported over a distance of approximately 2 m into the trapping potential of the high-field trap (shown in Fig.~\ref{fig:trap}(b)) with a kinetic energy of 43 eV. A gate electrode, driven by a fast high-voltage switch, is briefly opened to admit each incoming positron pulse and subsequently closed to prevent particle loss. A second electrode located downstream of the gate is biased to a lower potential to confine particles already stored in the trap.

During the initial injection phase, positrons are cooled exclusively by radiation (cyclotron cooling), with a characteristic time constant of about 160 ms in the 5 T magnetic field. 
%%% chat-gpt calculates 100 ms!  (it also says "cyclotron cooling" is more appropriate than "synchrotron cooling" in the non-relativistic regime :-)
As accumulation proceeds, the stored cold plasma contributes to the cooling of subsequently injected particles. To account for the increasing plasma space-charge potential and to limit injection-related acceleration at the beginning of the sequence, the depth of the trapping potential is progressively increased. After accumulating a sufficient number of pulses, the electrode voltages are adjusted to raise the bottom of the potential well to 500 V and to compress the plasma. The positron cloud is then released by rapidly switching the gate electrode to ground and is accelerated to 4 keV using a pulsed drift-tube. The operation of the high-field trap is similar to that described in Ref.~\cite{Blumer_2022}, but the potentials and timing were optimized for the new configuration.

\subsection{Record positron accumulation}
\label{sec:results}

\begin{figure}
    \centering
    \includegraphics[width=0.6\linewidth]{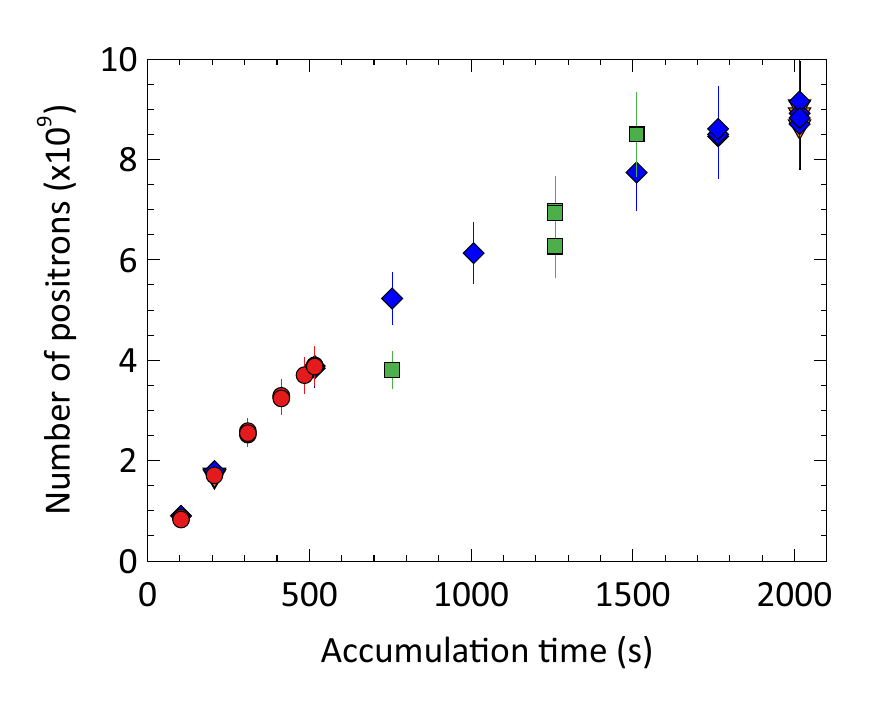}
    \caption{Positrons accumulated in the high-field trap versus time. The different symbols represent sets of measurements performed at different times.
    }  
    \label{fig:accumulation_figure}
\end{figure}

The number of positrons (as determined using the calibrated scintillators surrounding the setup) versus accumulation time is shown in Fig.~\ref{fig:accumulation_figure}. The uncertainty on the absolute positron number is dominated by the calibration of the CsI scintillators against the charge-sensitive preamplifier measurement and possible variations in the annihilation geometry. The total systematic uncertainty is estimated to be 10~\%. The number of trapped positrons increases nearly linearly up to almost $4\times10^{9}$.  Beyond that, the rate of accumulation decreases but the number of particles still increases, until $8.9\pm0.9\times10^{9}$. Measurements at the longest accumulation time of 2000~s were repeated seven times on two different days, resulting in a $0.15\times10^{9}$ standard deviation in positron number. Attempts to trap larger numbers of positrons led to almost complete losses. The depth of the potential well used to collect the highest number of positrons was almost 600~V and the estimated length of the positron plasma was approximately 300~mm. We could accumulate the $8.9 \times 10^9$ positrons in about 35 minutes, which is significantly higher, both in number and in accumulation rate, than the record value reported by ATRAP ($4 \times 10^9$  in 4.5 hours \cite{Fitzakerley_2016}).

The incorporation of the SiC remoderator significantly enhances accumulation in the high-field trap compared with the previously used conventional Surko-type configuration.  The accumulation rate is approximately 3.5 times higher than that achieved with the original configuration of the GBAR positron line, which reached a population of $1.4 \times 10^9$ positrons in 1100~s \cite{Blumer_2022}. Within the duration of the ELENA antiproton extraction and cooling cycle (typically 115~s), up to $6\times10^{8}$ positrons can be accumulated, including the time required for rotating wall compression, typically performed at a frequency of 6~MHz and an amplitude of 7.6~V$_{\mathrm{pp}}$.

\section{Summary}
We have demonstrated the accumulation of $(8.9\pm0.9)\times10^9$ positrons in the GBAR high-field Penning--Malmberg trap using a linac-based positron source coupled to a buffer-gas trap equipped with a movable SiC remoderator. The modified buffer-gas-trap configuration provides efficient capture and cooling of the pulsed positron beam, more than doubling the efficiency achieved with the previous Surko-type configuration. Under nominal operating conditions, more than $(6\pm0.6)\times10^8$ positrons can be accumulated within a single ELENA cycle. Longer accumulation sequences yield up to $(8.9\pm0.9)\times10^9$ positrons in less than 35 minutes, corresponding to an average accumulation rate of approximately $(4.5\pm0.5)\times10^6$~e$^+$/s. This represents more than twice the previously reported maximum trapped-positron population of  $4\times10^9$ positrons \cite{Fitzakerley_2016}, which was accumulated over 4.5 hours. Further improvements in the transport of these positron clouds will enhance positronium production and support the future synthesis of antihydrogen ions. The demonstrated trapping scheme may also be applicable to other experiments requiring intense, pulsed low-energy positron beams.

\section{Acknowledgments}
This work was carried out as part of a project co-financed by the Polish Minister of Education and Science under contract No. 2022/WK/11. This work is supported by: JSPS KAKENHI Grant-in-Aid for Scientific Research A 20H00150 and Fostering Joint International Research A 20KK0305 (Japan), SPHINX ANR-22-CE31-0019-01 and ANTION ANR-14-CE33-0008 (France), Programme National Gravitation, Références, Astronomie, Métrologie (PNGRAM), Institut de Physique and Noyaux et Particules, CNRS (France), the Swiss National Science Foundation (Switzerland) grants 197346, 232699 and 216673 and the ETH Zurich (Switzerland) grant ETH-46 17-1, the Swedish Research Council (VR) grants 2017-03822 and 2021-04005 and the following grants from Korea: IBSR016-Y1, IBS-R016-D1, UBSI Research Fund (No. 1.220116.01) of UNIST, NRF-2016R1A5A1013277, NRF- RS-2022-00143178, NRF2021R1A2C3010989, and NRF-2016R1A6A3A11932936. The GBAR collaboration is an International Research Network, supported by CNRS, France.

% old \bibliographystyle{unsrt}
%\bibliographystyle{iopart-num}
\bibliographystyle{iopart-num-title-doi}
%\bibliography{bibfile}

%\bibliographystyle{unsrtnat_ICPA-20}
%\bibliographystyle{iopart-num-long.bst}
%\bibliographystyle{unsrt.bst}
%apalike,unsrtnat
% old \bibliography{references}
\bibliography{references_IOP_manuscript}

\end{document}